\documentclass[preprint2]{aastex}

\slugcomment{To appear in Astronomical J., 45.}

\shorttitle{Dynamical Evolution of the TW Hya Association}
\shortauthors{de la Reza et al.}

\begin{document}


\title{Dynamical Evolution of the TW Hya Association}

\author{R. de la Reza\altaffilmark{1},  E. Jilinski \altaffilmark{1, 2}
and V. G. Ortega\altaffilmark{1}}

\altaffiltext{1}{Observat\'orio Nacional, Rua General Jos\'e
Cristino 77, S\~{a}o Cristov\~{a}o, 20921-400, Rio de Janeiro,
Brazil.}

\altaffiltext{2}{Pulkovo Observatory, Russian Academy of Science,
65, Pulkovo, 196140 St. Petersburg, Russia.}

\email{delareza@on.br,jilinski@on.br,vladimir@on.br}

\begin{abstract}
Using Galactic dynamics we have determined the age of the low mass
post-T Tauri stars TW Hya Association (TWA). To do so we applied
the method of Ortega et al.(2002, 2004) to five stars of the
association with Hipparcos measured distances (TWA 1, TWA 4,TWA
9,TWA 11,TWA 19).The method is based on the calculation of the
past 3D orbits of the stars. Of these stars only TWA 9 presents a
quite different orbit so that it does not appear to be a dynamical
member of TWA. The four remaining stars have a first maximum
orbits' confinement at the age of -8.3 $\pm{0.8}$ Myr which is
considered the dynamical age of TWA. This confinement fixes the
probable 3D forming region of TWA with a mean radius of 14.5 pc.
This region is related to the older subgroups of the Sco-Cen OB
association, Lower Centaurus Crux (LCC) and Upper Centaurus Lupus
(UCL), both with a mean age of about 18 Myr. This dynamical age of
TWA and that of the $\beta$ Pic Moving Group (BPMG) with an age of
11 Myr, also discussed here, introduce a more precise temporal
scale for studies of disks evolution and planetary formation
around some stars of these associations. Using the retraced orbit
of the runaway star HIP 82868 we examine the possibility that the
formation of TWA was triggered by a supernova (SN) explosion. It
is shown that for the four considered TWA stars, the expansion in
volume is a factor of five since their origin to the present
state. This is mainly due to the presently more distant star TWA
19.

\end{abstract}


\section{Introduction}

In a seminal paper, Herbig (1978) listed a few T Tauri stars
isolated from clouds; he proposed that they could be examples of
the long unknown post-T Tauri stars (PTTS). Among these was the K
type star TW Hya whose typical T Tauri nature was later disclosed
by Rucinski \& Krautter (1983). Afterwards, based on far-IR IRAS
sources, further T Tauri type stars were discovered in a radius of
about 5 degrees around TW Hya (de la Reza et al. 1989). At that
time, their relative high Galactic latitudes suggested that these
objects could be nearby stars. Rucinski (1992) confirmed the
isolation  of these stars from clouds and produced independent
arguments compatible with their youth. During the initial years of
the Pico dos Dias Survey (PDS), based on IRAS sources, new stars
were added to this group (Gregorio-Hetem et al. 1992). The studies
of this remarkable association have progressed with the facilities
of the Hipparcos (astrometry) and the ROSAT (x-rays)
observatories. Not only was it possible to confirm their proximity
(50 - 60 pc) but also to detect new members. This improvement was
the result of the efforts of several research groups; see for
instance, Kastner et al. 1997, Jensen, Cohen \& Neuh\"auser 1998,
Soderblom et al. 1998, Weeb et al. 1999, Sterzik et al. 1999,
Zuckerman et al. 2001a, Torres et al. 2003, Reid 2003, Song,
Zuckerman \& Bessell 2003, 2004. For a recent review see:
Zuckerman \& Song 2004.

In general, the main observational criteria or conditions to
establish the association status of a moving group as TWA, or of
others such as BPMG, are the following: a) similarity of their
space velocities characterizing the association as a probable
coeval moving group b) the same age of their members obtained by
means of pre-main sequence Hertzsprung-Russell (HR) evolutionary
diagrams c) the presence of spectroscopic features characterizing
their youth such as a relative strong Li absorption line and
H$\alpha$ lines with moderate emission/absorption or filled-in
properties.

In this work we use a different approach to determine the age and
the origin of TWA. This is a dynamical approach in which the 3D
orbits of the stars are retraced to find the orbits' confinement
under the action of a Galactic potential. This methodology has
already been used to study BPMG (Ortega et al. 2002, 2004) and the
$\epsilon$ and $\eta$ Chamaeleontis groups (Jilinski et al. 2005).
Two papers (Makarov, Gaume \& Andrievsky 2005, and Mamajek 2005)
were recently published, discussing the expansion of TWA. In the
following sections we shall comment on them and compare their
results with those obtained in the present work.

In Section 2 the methodology, the data and the dynamical results
of TWA are presented. Section 3 is devoted to the discussion of a
possible origin of TWA and finally discussion and conclusions are
the subject of Section 4.

\section{The Dynamical Evolution of TWA }

\subsection {The Method}

The stars of an unbound young stellar group develop their 3D
orbits and change their spatial velocities because of the action
of the general gravitational field of the Galaxy. The 3D orbits of
the stars are a fundamental element in the scenario we are
considering. The latter requires the use of a good, realistic mass
model of the gravitational Galactic potential. As stated in our
previous work (Ortega et al. 2002) the orbits integrations have
been performed using the Galactic potential by Hoogerverf, de
Bruijne \& de Zeeuw (2001). The parameters predicted by the model
are as follows: R$_\odot =$ 8.5 Kpc, V$_\odot =$ 219.8 km
s$^{-1}$, A $= 13.5$ km s$^{-1}$ kpc$^{-1}$ B $= -12.4$ km
s$^{-1}$ kpc$^{-1}$ and $\rho_0 $ $\sim 0.1$ M$_\odot$ pc$^{-3}$
where R$_\odot$ is the galactocentric distance of the LSR,
V$_\odot$ its rotation velocity, A, B the Oort rotation parameters
and $\rho_0 $ the local mass density. All these values are
consistent in relation to those obtained using data from
HIPPARCOS. Some authors (e.g., Ma\'iz-Apell\'aniz 2001, Makarov,
Ollin \&Teuben 2004) employ the perturbative, linear epicyclic
approximation of the equations of motion in which the vertical
displacement of the stars takes the form of harmonic oscillations
relative to the equatorial plane of the Galaxy. However, in view
of the fact that the phenomenon of 3D orbits confinement is very
sensitive to the "W" component of the space velocity vector, we
prefer to use the full system of the equations of motion. This was
also the approach we followed in all our papers mentioned above
where more details of the method are given.

\subsection {The Data for TWA}

The 3D trajectories of the stars of the group are integrated
backward in time starting from their initial, present positions
and velocities in a heliocentric Galactic oriented frame of
reference. Since the dynamical determination of the group's age
requires the best possible kinematic data we utilize here only
Hipparcos measured parallaxes. Of the classified TWA stars only
the following five fulfill this condition: HIP 53911 (TWA 1) with
distance of 56.4 pc, HIP 55505 (TWA 4) with - 46.7 pc, HIP 57589
(TWA 9) with - 50.3 pc, HIP 61498 (TWA 11) with - 67.1 pc and HIP
57524 (TWA 19) with - 103.9 pc. In our calculations we use the
spatial velocities of these stars determined by Reid (2003). These
were found by means of the use of Hipparcos astrometric data and
observed radial velocities which in general agree better than 1 km
s$^{-1}$ with those recently observed by Torres et al. (2003).
Some relatively small differences exist for stars TWA 11 and TWA
19 for which the values proposed by Torres et al. (2003) have the
largest uncertainties equal to $\pm 2.3$ and $\pm 3.8$ km s$^{-1}$
respectively. For TWA 11 Reid's radial velocity value is 3 km
s$^{-1}$ less than the mean value of Torres et al. (2003), whereas
TWA 19 Reid's value is within the mentioned uncertainty of this
star.

\subsection {The Dynamical age of TWA}

Our dynamical calculations were performed for the five TWA
classified stars mentioned above. Of these, only TWA 9 escapes
confinement if we take its distance as the mean Hipparcos value
(50.3 pc). This star was then eliminated as a dynamical member of
TWA. The remaining four stars confine well at – 8.3 Myr. In Figure
1 (a and b) we present the results of the past 3D orbits evolution
of the four TWA here considered stars. The galactic (X,Y,Z)
coordinates are positive in the direction of the Galactic center,
of Galactic rotation and the North Galactic pole respectively.
(X,Y) is the plane through the Sun and (Z,Y) is orthogonal to this
plane. In Figure~2 is shown the variation with time of the 3D
radius of the spatial distribution of the four TWA stars relative
to the group center. The minimum radius of 14.5 pc represents the
maximum 3D orbit confinement attained at the age of $-8.3$ Myr.
This radius variation shows that the group formed by these four
stars has expanded in volume, from its initial state up to its
present state with a radius of 24 pc, by a factor of five. The
internal errors in the present work were estimated by realizing
1000 Monte Carlo simulations for each star. This was done by
randomly choosing initial values of the space velocities from a
Gaussian distribution with mean values equal to the observed ones
and a dispersion of 1 km s$^{-1}$. The mean radius for the
obtained forming region of TWA is of 14 pc. The cumulative effect
of the uncertainties for each orbit represents, at the considered
birthplace of TWA, an external shell with a thickness of about 2
pc corresponding to a dynamical age error of $\sim$ 0.8 Myr.
Typical uncertainties estimated by this method for different past
epochs can be seen in Jilinski et al. (2005).

\begin{figure}
\epsscale{1.5} \plottwo{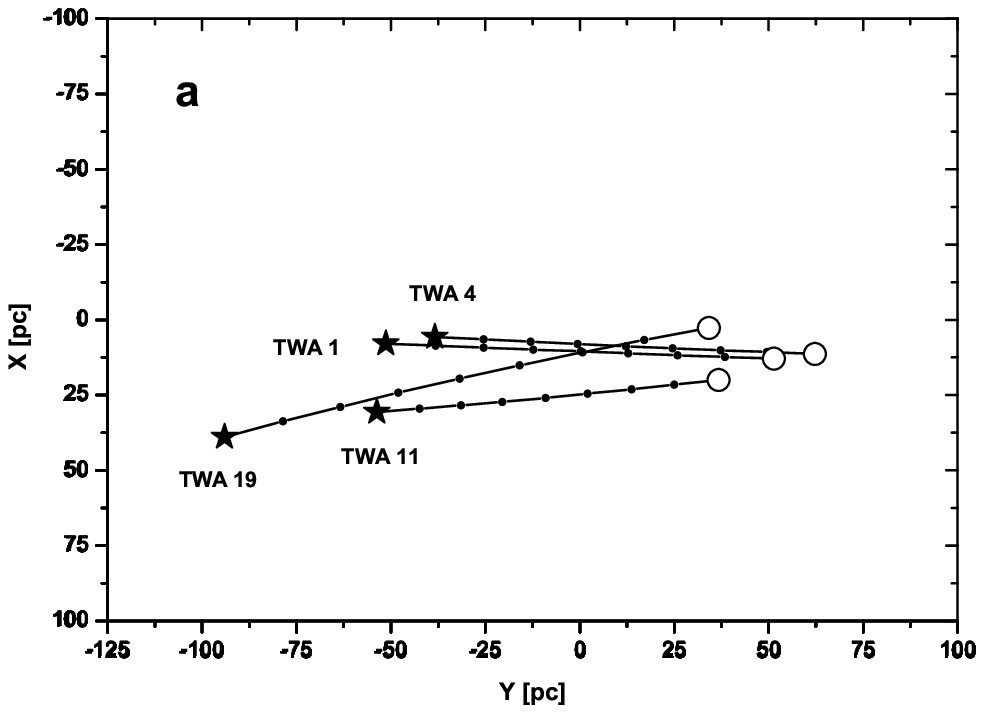}{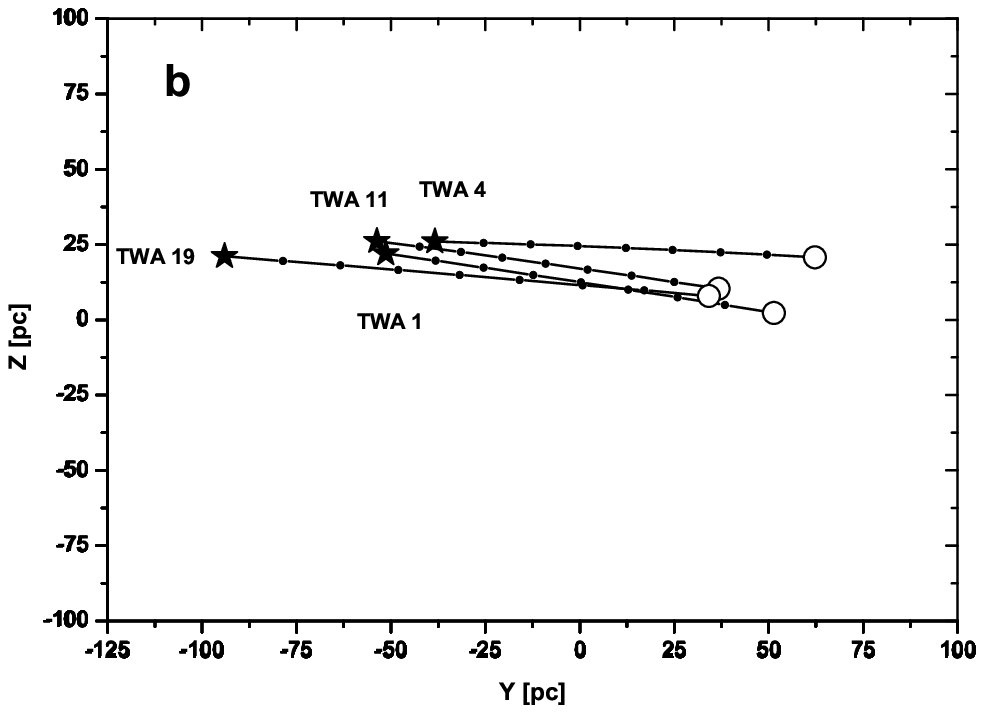} \caption{Individual
orbits of TWA members in the (X,Y) plane (\textbf{a}) and in the
(Z,Y) plane (\textbf{b}). Each two point interval corresponds to 1
Myr. The size of the circle is proportional to the uncertainty in
the calculated star position at -8.3 Myr. \label{fig1}}
\end{figure}

\begin{figure}
\epsscale{0.8} \plotone{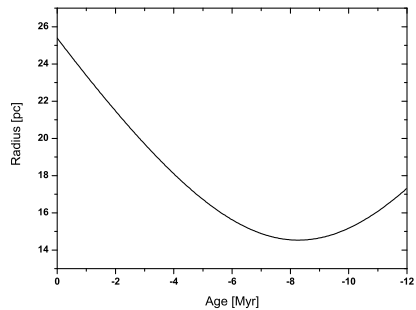} \caption{Variation with time of
the 3D radius of the distribution of stars TWA~1, TWA~4, TWA~11
and TWA~19. The dynamical age (-8.3 Myr) is determined by the
maximum 3D confinement at the minimum of the curve. \label{fig2}}
\end{figure}

\section{The Birthplace of TWA}

In Figures 3(a and b) we schematically present the dynamical
evolution of the centroids of TWA, LCC and UCL in (X,Y,Z). The
importance of considering 3D can be seen from the differences in
the trajectories of TWA and BPMG: they appear to be fairly similar
in the (X,Y) plane however, differences appear in the Z direction
(see Figures 4 a and b).

We note that the trajectories of LCC and UCL represent the
centroid evolution of hot stars of these subgroups (see Ortega et
al. 2004). Differently from the cases of TWA and BPMG, no
confinement considerations were done for LCC and UCL. Their past
sizes remain then undefined. For practical reasons, we maintained
in the past their approximate, presently observed sizes. In any
case the confined region for TWA, which we consider to be the
birthplace of TWA, appears to be related in time and space to both
LCC and UCL subgroups. Similar relation between TWA, LCC and UCL
subgroups was considered by Mamajek, Lawson \& Feigelson (2000) on
the basis of a linear extrapolation of velocity vectors in the
Galactic plane.

In Ortega et al. (2004) we discussed the possibility that a single
SN event could eventually have triggered the formation of BPMG 11
Myr ago. Could a similar event be involved also in the formation
of TWA? One way to obtain information about this consists in
finding a presently observed hot runaway star that could be the
remanescent of a binary star, one of the components of which
exploded as a SN. Similarly as in Ortega et al. (2004), using the
list of runaway stars in Hoogerwerf, de Bruijne \& de Zeeuw
(2001), we found that the Be star HIP 82868 could be an
interesting candidate for taking part in the triggering of TWA
formation. We should note however, that the Hipparcos distance and
the radial velocity of HIP 82868 have uncertainties of $\sim 20\%$
and $\sim 25\%$ respectively. These uncertainties may produce
errors in the past orbit of this star in the sense of either
approaching it, or removing it from the TWA forming region. In the
former case the triggering mechanism would win in efficiency while
in the latter it would lose it almost completely. In any case, it
seems advisable to take into account this conjecture as a possible
mechanism of triggering TWA. Hereafter, we will consider only the
mean values of these two parameters. As in Hoogerwerf et al.
(2001)we also found that this runaway star crossed the open
cluster IC 2602 some 6 Myr ago. Nevertheless, because this cluster
is older than 25 Myr, HIP 82868 can hardly be considered to be
related to this cluster. As a matter of fact, several runaway
stars studied by Hoogerwerf et al. (2001) crossed one or two
different structures during their past lives. As far as the
cluster IC 2602 is concerned, we cannot exclude the possibility
that some gravitational effect produced by this cluster could have
somewhat modified the orbit of HIP 82868, but this will not be
considered here.
 Using the same methodology as described in Section 2,
it was found that the 3D orbit of this star was, at -9 Myr, about
90 pc away from the birthplace center of TWA (see Figures 4 and
5). That means that the shock front travelled this distance in
about 1 Myr before reaching the region of the TWA formation.

%\clearpage

\begin{figure}
\epsscale{1.5} \plottwo{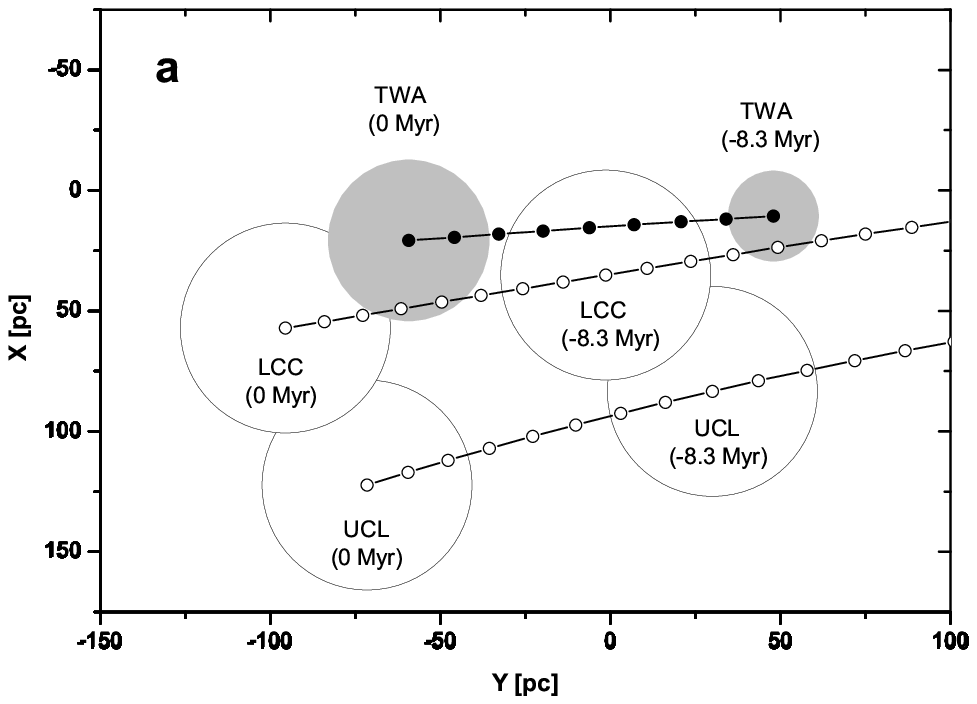}{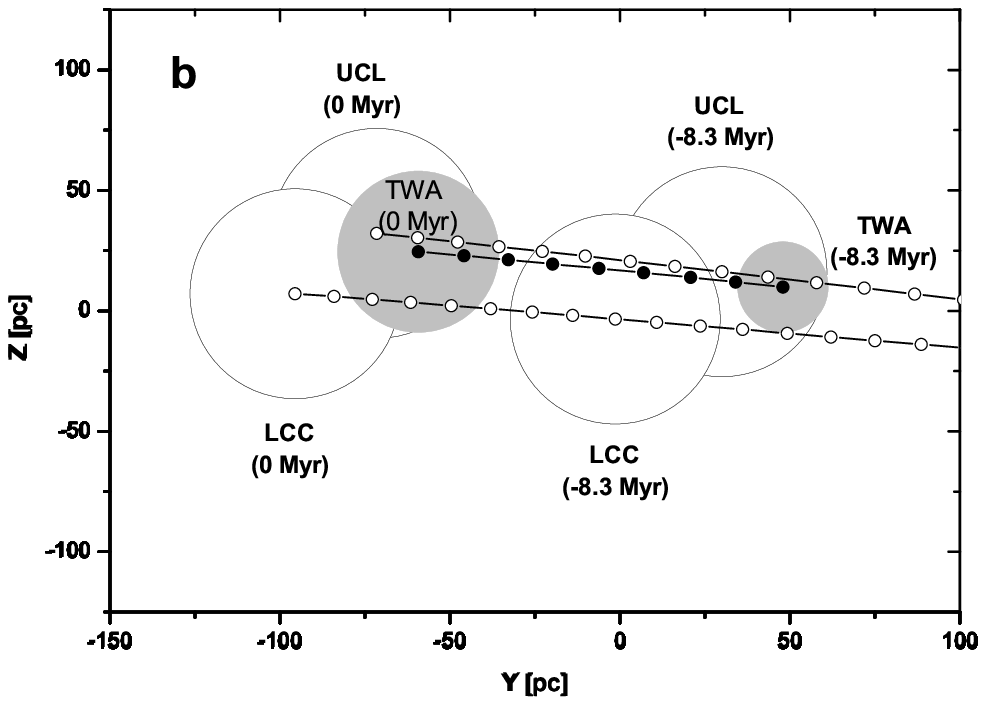} \caption{Past orbits
evolution of LCC, UCL and TWA groups. The orbit of the kinematical
center of TWA is shown until the age of 8.3 Myr. Each two point
interval corresponds to the motion during 1 Myr. \label{fig2}}
\end{figure}

\begin{figure}
\epsscale{1.5} \plottwo{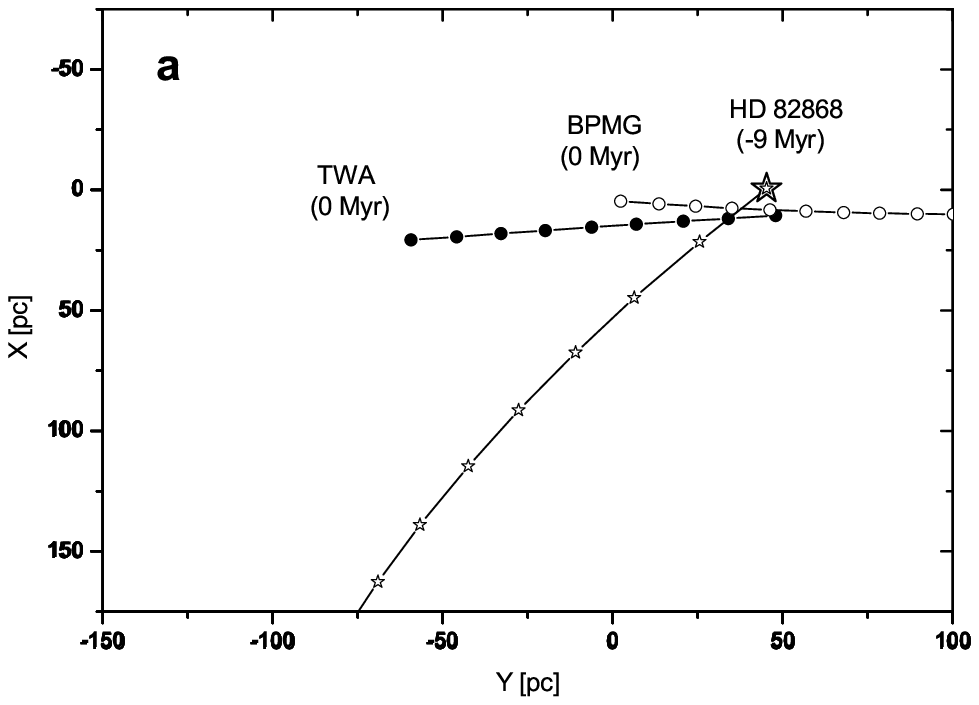}{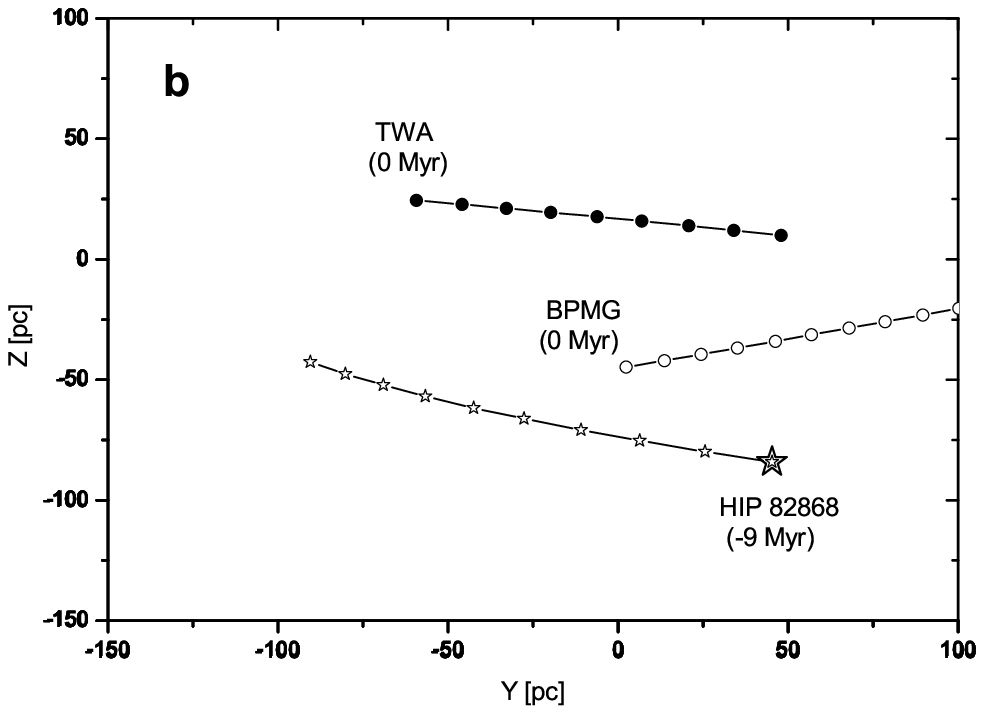} \caption{Past orbits
evolution of the kinematic centers of BPMG, TWA and the runaway Be
star HIP 82868 in the same planes as in the Fig. 1. The position
of the proposed SN explosion 9 Myr ago is shown as a big star.
Each two point-interval corresponds to the motion during 1 Myr.
\label{fig3}}
\end{figure}

%\clearpage

Approximate and reasonable physical conditions can be found for
this scenario considering Sedov's relations R =
1.17(E$_o/{\rho}_o)^{1/5}$ t$_{SN}^{2/5}$ and  V$_{SN} =
(2/5)$(R/t$_{SN})$(see, e.g. Boers \& Deeming 1984) where R is the
radius of the shock front moving with a terminal velocity V$_{SN}$
and t$_{SN}$ being the time elapsed since the explosion. $\rho_o$
is the interstellar density outside the front. For our case, R =
90 pc and t$_{SN}$ = 1 Myr giving V$_{SN}$ = 36 km s$^{-1}$. We
can then estimate the energy (E$_o$) of the SN for different
values of $\rho_o$. Choosing a density value corresponding to one
H atom 1 cm$^{-3}$ we obtain E$_o$ = 1.3$\times$ 10$^{51}$ ergs ,
consistent with the canonical value of 10$^{51}$ value expected
for stars with masses of 15 and 25 $M_{\odot}$ or more and with
solar abundances (Woosley, Heger \& Weaver, 2002). Somewhat
smaller but equally realistic values for E$_o$, can be obtained
admitting smaller interstellar densities resulting from the action
of the strong stellar winds of pre-supernova hot OB stars.

We notice that as shown in Figures 4 and 5, the 90 pc 3D distance
of the proposed SN relative to the center of the birthplace of TWA
is due mainly to the "Z" coordinate. While the TWA birthplace was
at the (X,Y) plane, the SN event occurred below this plane (see
Figures 4 (a and b)). It is interesting to note that the resulting
dynamical evolution of TWA from its origin to its place today
follows, even with a small angle, the correct increasing "Z"
direction. This is to be expected if the shock front of the SN not
only took part in triggering the star formation of TWA, but also
produced an extra impulse to the natal cloud making the spatial
motions of the unbound stars of the association to follow
approximately in the direction of the shock front. We note that
the explosion of the SN is considered here to be isotropic and its
initial shock front velocity several orders of magnitude larger
than the initial velocity of the runaway star. Under these
conditions the motion of the shock wave and the runaway star are
practically independent.

It is also interesting to note that at the proposed epoch of the
SN explosion at -9 Myr, the two stars $\beta$ Pic and AU Mic
pertaining to BPMG with an age of 11 Myr and both with observable
disks were far apart from the SN at distances of about 96 pc and
120 pc respectively. This is also the case at the age of -8 Myr
during the formation of TWA, when the distances of $\beta$ Pic and
AU Mic to the birthplace center of TWA were respectively about 33
pc and 47 pc. As the shockfront is isotropic, it will attain these
stars, specially $\beta$ Pic. If our scenario of TWA formation is
true the disks of these stars, observed today, have somehow been
preserved. One possibility of survival could be due to their very
small cross sections as compared with the parent cloud of TWA. In
Figures 5 (a and b) we show the positions of the stars $\beta$ Pic
and AU Mic together with the birthplace of TWA at the moment of
TWA formation and also the position of the runaway star HIP 82868
at the moment of the SN event (-9.0 Myr).The positions of $\beta$
Pic and AU Mic at -9.0 Myr are also shown.

%\clearpage

\begin{figure}
\epsscale{1.5} \plottwo{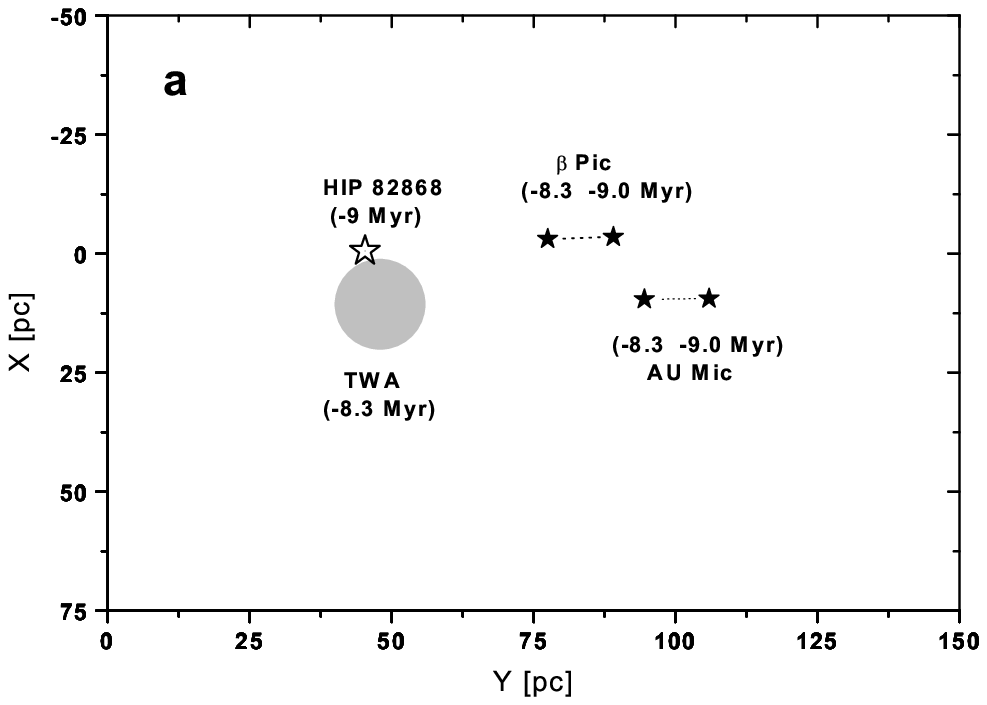}{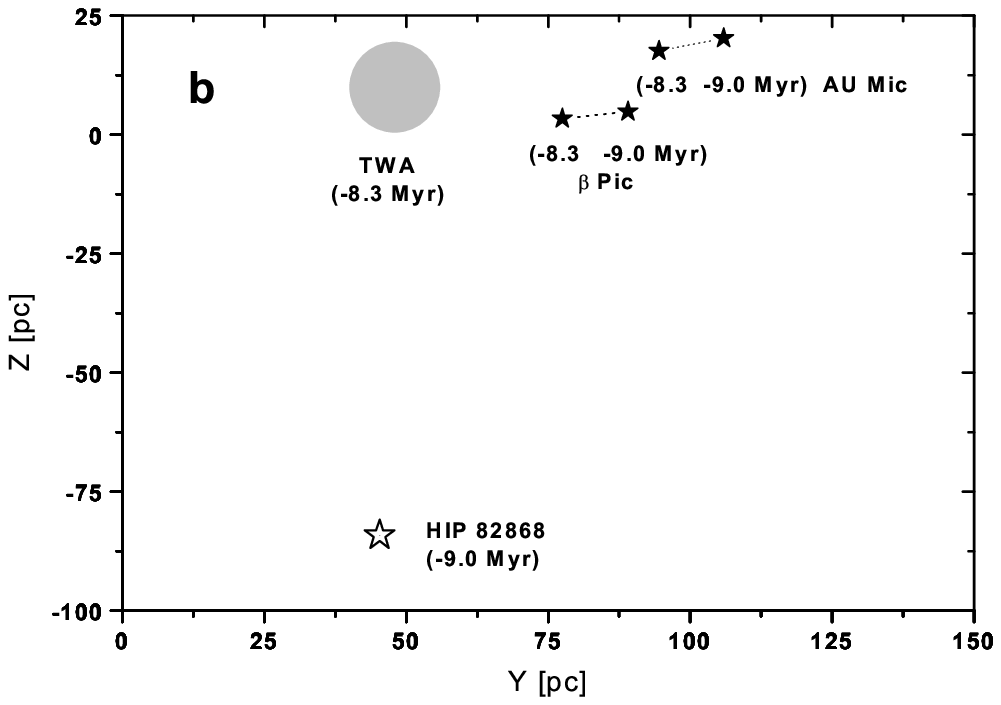} \caption{Mutual
positions of TWA, $\beta$ Pictoris star and AU Mic at the moment
of TWA formation in the (X,Y) and (Y,Z) planes. The position of
proposed SN explosion which took place 9 Myr ago is presented as a
big white star. The positions of stars the $\beta $ Pic and AU Mic
at 9 Myr ago are also shown. \label{fig4}}
\end{figure}

%\clearpage

\section{Discussion and Conclusions}

At least for quite young stellar groups, the method of retracing
the 3D orbits of their stars, using a model of the general
Galactic potential allows one to investigate important questions
concerning those groups such as finding probable birthplaces,
dynamical ages and identifying true group members. In this work,
we have explored the dynamical evolution of TWA by considering
five TWA classified stars for which Hipparcos distances are known.
Of these stars only TWA 9 showed a past orbit different from the
rest. This star is then not considered to be a dynamical member of
TWA. All the other stars TWA 1, TWA 4, TWA 11 and TWA 19 confine
at – 8.3 $\pm$ 0.8 Myr in a spatial region with an average radius
of 14.5 pc. One important consequence, mainly due to the present
distant star TWA 19, is that the volume expansion of these four
stars is of a mean factor of five. This result can be obtained by
comparing the present time volume occupied by these stars with
their volume at the origin $\sim$ 8.3 Myr ago. This region appears
to be related to past positions of both LCC and UCL subgroups of
the Sco-Cen OB Association. We remark that in this study we cannot
establish the membership of other possible members of TWA
suggested in the literature because they all lack Hipparcos
distances. Recently, Makarov et al. (2005) also studied the
dynamical expansion of TWA using the epicycle approximation for
the equations of motion. Their sample is not completely the same
as ours. Even though they considered the stars TWA 1, TWA 4 and
TWA 11 which are in our sample, they introduced three possible
members: HD 139084 (HIP 76629), a member of BPMG according to
Zuckerman et al. (2001b), HD 220476 and the star Vega. With this
sample they find an expansion age of 4.7 $\pm$ 0.6 Myr smaller
than ours. By taking into acount the discrepancy between this
small expansion age and the isochrone age of TWA, which is between
8 and 10 Myr as found in the literature, Makarov et al. introduced
two episodes during the formation to explain the origin of TWA. In
a first one, the stars were formed and the parental cloud was
maintained intact during some Myrs. In a second one, at –4.7 Myr
some external agent removed the gas of the natal cloud, probably
by a collision with an another cloud, allowing in this way the
stars to begin their motions as independent systems. In our
scenario, it is proposed that a possible violent formation of TWA
by a SN event could, in favorable conditions, not only form the
stars of TWA 8.3 Myr ago, but also disperse the remaining original
gas leaving an unbound stellar system since its origin. So the
difference of the results of Makarov et al. work and ours can be
explained by different adopted methodologies and samples of stars.
Mamajek (2005) applies the linear expansion method of Blaauw
(1956), which makes use of the proper motions, to a sample of 23
stars (without TWA 19AB) finding only weak evidence for the
expansion. He notes, however, that his sample is consistent with a
lower limit of 10 Myr for an expansion age.

The fact that TWA 19AB belongs to this young coeval moving group
has some other consequences , apart from determining the 3D
expansion of TWA. Lawson \& Crause (2005) propose that TWA
consists of two groups, one nearby, younger with an age of about
10 Myr containing stars with longer rotational periods and an
older (17 Myr) distant one, associated with LCC and whose stars
have shorter rotational periods. The rotational period differences
between the stars of those groups are thought to be due to an
spin-up process taking place during that age interval. The
physical binary TWA19AB, for which they did not measure the
period, appears to play an important role as calibrator, because
it is the only star in the distant group for which an Hipparcos
distance is known. In order to compatibilize the HR ages of TWA
19A (17 Myr) and of TWA 19B (5 Myr) they adopt an ad-hoc
hypothesis suggesting that TWA 19B is by itself a binary star with
equal luminosity components. Our results indicate that TWA 19AB
has a mean age of 8.3 Myr and that this ad-hoc hypothesis is then
unnecessary.

If the formation of TWA has been assisted by a SN, we found a
possible past location of this event by retracing the 3D orbit of
the runaway Be star HIP 83868. We discussed and examined plausible
physical conditions supporting this scenario. An isotropic
evolution of the SN remnant is also expected to reach the stars AU
Mic and especially $\beta$ Pic both belonging to BPMG. Then, if
our scenario of TWA formation is true we conclude that some effect
(perhaps tiny cross sections) has contributed to preserve the
disks from destruction. Eventual destruction of stellar disks by
the action of SN remnants (as far as we know) had not been
considered in the literature.

TWA was always considered in the literature as the best
"laboratory" for studies of protoplanetary disks and early
planetary formation. In fact, some member stars as TW\,Hya=HIP
53911 (TWA 1), Hen 3.600A (TWA3 A), HIP 55505 (TWA 4) have disks
of the T Tauri type and a debris disk of the annular type in the
star HIP 61498 (TWA 11A). We recall also that BPMG, with a
dynamical age around 11 Myr (Ortega et al. 2002, 2004), has as
mentioned before, two stars with debris disks of the "$\beta$ Pic
type": the proper $\beta$ Pic star and the star AU Mic. We can
then tentatively suggest that the age of  8.3 Myr marks the onset
of debris disks formation, at least for young stars in
associations. Presently it is not yet clear which is the typical
age characterizing the end of the T Tauri type disks. Recently,
Torres et al. (2006) concluded that the spectroscopic binary
classical T tauri star V4046 Sgr, containing an important
circumbinary disk (Quast et al. 2000, Stempels \& Gahm 2004), is
probably a member of BPMG.

Concerning planets, a giant one with a mass of 5 M$_J$ has been
detected recently near the brown dwarf star 2MASS WJ
1207334-393254 (Chauvin et al. 2005). This star was proposed by
Gizis (2002) to belong to TWA  which was confirmed by  Mamajek
(2005) on the basis of its kinematics. The formation mechanism of
this planet is not clear however; it may have been formed either
in the disk of the brown dwarf or together with this star.
Whatever the involved mechanism, the age of TWA fixes a time scale
of this mechanism. It also indicates that any initial planet
building process, as that observed in the disk of the star TW Hya
(Wilner et al. 2005), could have been initiated before 8 Myr.

For BPMG there are very strong indications of the presence of
hidden planets in the debris disks of $\beta$ Pic and AU Mic stars
(see for example Liu, 2004). Considering that the dynamical age of
BPMG is 11 Myr, we can suggest that these planets, if present,
were formed before or near this age.

\acknowledgments

We thank the referee for considerations that improve the
presentation of this paper. EGJ thanks MCT Brazil for financial
support.

\end{document}